\documentclass[12pt]{article} 

\usepackage{latexsym} 
\usepackage{amssymb}  
\usepackage{epsfig}       

\newcommand{\al}{\alpha}
\newcommand{\be}{\beta}

\newcommand{\de}{\delta}

\newcommand{\la}{\lambda}

\newcommand{\del}{\nabla}

\renewcommand{\th}{\theta}   

\newcommand{\p}{\partial}
\newcommand{\<}{\langle} 
\renewcommand{\>}{\rangle} 

\newcommand{\txt}{\textstyle}
\newcommand{\dsp}{\displaystyle}

\newcommand\eqn[1]{(\ref{#1})}      
\newcommand{\e}{ {\rm e} }
\newcommand{\beq}{\begin{equation}}
\newcommand{\eeq}{\end{equation}}
\newcommand{\ba}{\begin{array}}
\newcommand{\bea}{\begin{eqnarray}}
\newcommand{\ea}{\end{array}}
\newcommand{\eea}{\end{eqnarray}}

\newcommand\comment[1]{ \hbox{[{\it Comment suppressed here.}\/]} }
\newcommand\hide[1]{}

\newcommand{\tr}{\hbox{tr}}

\newcommand{\skipover}[1]{}

\newcommand{\half} {{\txt {1\over 2}}}
\newcommand{\third}{{\txt {1\over 3}}}
\newcommand{\twothirds}{{\txt {2\over 3}}}

\def\phm{\phantom{-}}

\makeatletter 

\def\appendix{\par                              
    \setcounter{section}{0}                     
    \setcounter{subsection}{0}
    \renewcommand{\theequation}{\Alph{section}.\arabic{equation}}
    \renewcommand{\thesection}{Appendix \Alph{section}
                \setcounter{equation}{0}  } 
}

\def\applabel#1{\@bsphack
  \protected@write\@auxout{}%
         {\string\newlabel{#1}{{\Alph{section}}{\thepage}}}%
  \@esphack}

\def\section{
\setcounter{equation}{0}        
\@startsection {section}{1}{\z@}{-3.5ex plus -1ex minus 
 -.2ex}{2.3ex plus .2ex}{\large\bf}}
\renewcommand{\theequation}{\arabic{section}.\arabic{equation}}

\def\subsection{\@startsection{subsection}{2}{\z@}{-3.25ex plus -1ex minus 
 -.2ex}{1.5ex plus .2ex}{\normalsize\bf}}

\def\subsubsection{\@startsection{subsubsection}{3}{\z@}{-3.25ex plus
 -1ex minus -.2ex}{1.5ex plus .2ex}{\normalsize}}

\makeatother   

\newsavebox{\eqlabel}

\makeatletter  
\newlength{\numblen}
\newsavebox{\eqnumb}
\def\@eqnnum{\savebox{\eqnumb}{\rm (\theequation)}%
\settowidth{\numblen}{\usebox{\eqnumb}}%
\makebox[\numblen][l]{\usebox{\eqnumb}~~~\usebox{\eqlabel}}}
\makeatother   

\newenvironment{equationwithlabel}[1]{ %
  \begin{equation}\label{#1} }{\end{equation}} 
\newcommand{\beql}[1]{\begin{equationwithlabel}{#1}}
\newcommand{\eeql}{\end{equationwithlabel}}

\newcommand{\Y}{T_8}
\newcommand{\R}{{\tilde Q}}
\newcommand{\Ar}{A^\R}
\newcommand{\Br}{B^\R}  
\renewcommand{\S}{X}
\newcommand{\As}{A^\S}
\newcommand{\Bs}{B^\S}  
\newcommand{\curl}{{\rm curl}}
\renewcommand{\div}{{\rm div}}
\newcommand{\MeV}{{\rm MeV}} 
\newcommand\colvec[2]{\left(\ba{@{\,}c@{\,}}#1\\#2\ea\right)}

\begin{document}

\title{\bf Magnetic fields within color superconducting neutron star cores}

\author{
	Mark Alford, J\"urgen Berges and Krishna Rajagopal \\[0.5ex]
{\normalsize Center for Theoretical Physics}\\
{\normalsize Massachusetts Institute of Technology}\\
{\normalsize Cambridge, MA 02139 }
}

\newcommand{\preprintno}{
  \normalsize MIT-CTP-2906 
}

\date{\today \\[1ex] \preprintno}

\begin{titlepage}
\maketitle
\def\thepage{}          

\begin{abstract}
We discuss the Meissner effect for a color superconductor formed by
cold dense quark matter. Though color and ordinary electromagnetism
are broken in a color superconductor, there is a linear combination of
the photon and a gluon that remains massless. Consequently, a color
superconducting region may be penetrated by an external magnetic
field. We show that at most a small fraction of the magnetic field is
expelled, and if the screening distance is the smallest length scale
in the problem there is no expulsion at all.
We calculate the behavior of the
magnetic field for a spherical geometry relevant for compact stars.
If a neutron star contains a quark matter core, this core 
is a color superconductor. Our results demonstrate that such
cores admit magnetic fields without restricting them to 
quantized flux tubes.  Such magnetic fields within color
superconducting neutron star cores are stable on time scales
longer than the age of the universe, even if
the spin period of the neutron star is changing.
\end{abstract}

\end{titlepage}

\renewcommand{\thepage}{\arabic{page}}


\section{Introduction}

Conventional superconductors result from a condensate of
Cooper pairs of electrons.  Because the Cooper pairs have nonzero
electric charge, electromagnetic gauge invariance
is spontaneously broken: the photon gets a mass and 
weak magnetic fields are expelled by
the Meissner effect.  
In this paper we show that the same is not in
general true for the color superconducting state that is formed by
cold dense 
quarks \cite{Barrois,BailinLove,ARW2,RappEtc,BergesRajagopal,ARW3,Reviews}, 
even though here also the Cooper pairs have nonzero 
electric charge. The reason is that a color
superconductor is not quite an electric superconductor: it makes the
gluons massive (there is a {\em color} Meissner effect) but does not
simply make the photon massive.  Rather, one linear combination of the
photon and a gluon becomes massive, but the orthogonal combination
remains massless.  Thus a region of color superconductor can allow
itself to be penetrated by the component of an external magnetic field
that corresponds to the unbroken generator.  As we will see below,
in the limit in which the screening length is the shortest
length scale in the problem, the magnetic field within
the color superconductor has the same strength as the
applied external magnetic field.  There is {\it no} Meissner effect.
Though the interior field is not diminished in strength, it is
``rotated'' relative to the external field: it is 
associated with the $U(1)$ symmetry which is unbroken 
within the superconductor, not with the $U(1)$ of ordinary electromagnetism.
If the penetration length 
is not smaller than the
thickness of the boundary 
of the color superconducting region (to be defined below), 
there is a partial Meissner effect. In this case, the strength
of the field which penetrates the superconductor 
depends on details of the geometry, the
relative sizes of the screening length and the boundary thickness, and
the relative strengths of electromagnetism and the color force.

The most likely place to find superconducting quark phases in nature
is in the core of neutron stars. 
If neutron stars achieve sufficient central densities
that they feature quark matter cores, these cores
must be color superconductors: because there are attractive
interactions between pairs of quarks which are antisymmetric
in color, the quark Fermi surfaces in cold dense quark
matter are unstable to the formation of a condensate
of quark Cooper pairs.
Present theoretical methods are not sufficiently accurate
to determine the density above which a quark matter description
becomes appropriate, and thus cannot answer the question of whether 
quark matter, and hence color superconductivity, occurs
in the cores of neutron stars.  What theory {\it can} do
is analyze the physical properties of dense quark matter
and, eventually,
make predictions for neutron star phenomenology,
thus allowing astrophysical observation to settle the question.

There are a number of avenues 
which may allow observations of neutron stars to answer 
questions about the presence or absence of color
superconductivity. (Examples we do not pursue in this paper 
include analysis of cooling
by neutrino emission \cite{Blaschke} and analysis
of r-mode oscillations \cite{Madsen}.)
Since neutron stars have high
magnetic fields ($10^8$ to $10^{13.5}$ Gauss in typical
pulsars \cite{Bhattacharya}; perhaps as high as
$10^{16}$ Gauss in magnetars \cite{ThompsonDuncan}) 
one prerequisite to making contact with neutron star phenomenology is
to ask how the presence of a
superconducting core would affect the magnetic field.
We will see in this paper that the strength of the magnetic field
within the superconducting core is hardly reduced, and the magnetic flux
within the color superconductor is not restricted to quantized
flux tubes. The latter constitutes a qualitative difference
between conventional neutron stars and those with quark matter
cores. We will see in Section 5 that, unlike in conventional
neutron stars, the magnetic field within a color superconducting
core does not vary even if the spin period of the neutron
star is changing.

\subsection{A fiducial example}

To give the reader some
sense for typical scales in the problem, we now
describe a fiducial example, which we will use in particular
in Sect.~\ref{sec:outlook}.  The numbers in this 
paragraph make the crude assumption that the
quarks are noninteracting fermions,
 and so should certainly not be construed  as precise.
Consider quark matter with 
quark chemical potential $\mu=400~\MeV$, made of
massless up and down quarks and strange quarks
with mass $M_s=200~\MeV$.  ($M_s$ is a density
dependent effective mass; this adds to the uncertainty
in its value.)  If the strange quark were massless,
quark matter consisting of equal parts $u$, $d$ and $s$
would be electrically neutral.  In our fiducial example,
on the other hand, electric neutrality requires a nonzero
density of electrons, with chemical potential $\mu_e=24~\MeV$.
Charge neutrality combined with the requirement that the
weak interactions are in equilibrium determine all the
chemical potentials and Fermi momenta:
\beq
\ba{rcrcl@{\qquad}rcl}
\mu_u &=& \mu - \frac{2}{3} \mu_e &=& 384~\MeV,
   & p_F^u  &=&  \mu_u, \\[2ex]
\mu_d &=& \mu + \frac{1}{3} \mu_e &=& 408~\MeV,
   & p_F^d  &=&  \mu_d, \\[2ex]
\mu_s &=& \mu + \frac{1}{3} \mu_e &=& 408~\MeV, 
   & p_F^s &=& \sqrt{\mu_s^2-M_s^2} = 356~\MeV, \\[2ex]
 && \mu_e &=& 24~\MeV,  & p_F^e &=& \mu_e~.
\ea
\label{fiducial}
\eeq
The baryon number density 
$\rho_B = (1/3\pi^2) [(p_F^u)^3 +(p_F^d)^3 + (p_F^s)^3]$ 
is 5 times nuclear matter density. 

A variety of estimates suggest that the gaps at the Fermi surfaces
resulting from 
quark-quark pairing are about $\Delta\sim 20-100~\MeV$,
resulting in critical temperatures above which the color
superconductivity is destroyed which are about $T_c\sim 10-50~\MeV$.
Neutron stars have temperatures $T\ll T_c$, and 
if they have quark matter cores these cores are certainly
in the superconducting phase.  Similar  estimates for $\Delta$ and $T_c$
are arrived at either by using phenomenological models, with
parameters normalized to give reasonable vacuum 
physics \cite{ARW2,RappEtc,BergesRajagopal,ARW3,CarterDiakonov,RappEtc2,Hsu1,SW0}
or by using weak coupling 
methods \cite{PisarskiRischke,Son,Hong,HMSW,SW3,PR,rockefeller,Hsu2}, 
valid for $\mu\rightarrow\infty$
where the QCD coupling $g(\mu)$ does become weak.  Neither strategy
can be trusted quantitatively for $\mu\sim  400~\MeV$, where
$g(\mu)\sim 3$, but it is pleasing that both strategies agree qualitatively.

\subsection{The CFL and 2SC phases of quark matter}

There are, in fact, two different superconducting phases
possible, which have very different symmetry properties.
If $\Delta$ is large compared to the differences among
the three Fermi momenta in (\ref{fiducial}),
$\langle ud \rangle$, $\langle us \rangle$ and $\langle ds \rangle$
condensates all form.  Chiral symmetry is broken by 
color-flavor locking \cite{ARW3} in this phase. 
This is the favored phase at large $\mu$ where
the differences between Fermi momenta decrease \cite{ABR,SW2}.
This CFL phase has the same symmetries as baryonic matter which
is itself sufficiently dense that the hyperon and nucleon
densities are all similar, and there need not be a
phase boundary between CFL matter and baryonic 
matter \cite{SchaeferWilczek,ABR,SW2}.
The properties of the CFL phase have been further
investigated in Refs. \cite{gapless,Gatto}.

Now, imagine starting with CFL matter at very
large $\mu$ and
reducing $\mu$. Because of the nonvanishing strange quark mass,
as $\mu$ decreases
the differences among the Fermi momenta
increase.  It may happen that before $\mu$ has decreased
so far that a quark matter description ceases to be valid,
one may find a phase of quark matter in which only
two flavors ($u$ and $d$) and two colors (chosen spontaneously)
of quarks pair.
Chiral symmetry is restored
in this two-flavor superconductivity (2SC) phase, which is also
the phase which arises in QCD with no strange quarks at 
all \cite{Barrois,BailinLove,ARW2,RappEtc,BergesRajagopal}.
Because nature chooses a strange quark which can neither
be treated as very light nor as very heavy, present
theoretical analyses are not precise enough to determine
whether quark matter at densities typical of neutron
star interiors is in
the CFL phase, or whether in this range of $\mu$
the 2SC phase is favored.  
For example, in (\ref{fiducial}) the splitting between
$d$ and $s$ Fermi momenta is $\sim 50$ MeV, of
the order of typical gaps. Current theoretical
methods are not reliable enough to determine whether 
the quark-quark interactions in QCD are 
strong enough to generate a $\langle ds \rangle$
condensate larger than $50$ MeV, as required in the CFL phase, or
are somewhat weaker, admitting only the 2SC pairing.  

In discussing conventional superconductors in a magnetic field,
one normally begins by distinguishing between Type I and
Type II superconductors. Color superconductors are Type I 
at asymptotically high densities \cite{HMSW,SW3,Hsu2}, but they
may be Type I or Type II at neutron star densities.
However, this distinction will
not be of importance.  First of all, the thermodynamic critical field
$H_c$ required to destroy the color superconductivity is
on the order of $10^{18}$ Gauss. 
If color superconductors
exhibited the conventional Meissner effect, whether
or not they were Type I or Type II they would
simply exclude neutron star magnetic fields,
which are in fact weak for our purposes.
Second of all, we shall see that the presence of
an unbroken rotated electromagnetism changes the
story completely.\footnote{
See \cite{BailinLove,Blaschke2} for early attempts to analyze
magnetic fields in color superconductors. These authors
neglect the existence of the unbroken rotated electromagnetism.
}
In the next three sections, 
we solve the problem of how a sphere of
color superconducting matter, which can admit
a rotated magnetic field, responds to an applied
ordinary magnetic field.  In the final section,
we discuss the consequences of our findings for
neutron stars.

\section{Rotated electromagnetism}

\subsection{The new photon}
The fundamental fields in the theory are the quarks $\psi^\al_i$ 
(with color index $\alpha$ and flavor index $i$)
and the gauge fields: 
the photon $A_\mu$ and the gluons $G^n_\mu$, $n=1\ldots 8$.
At low temperature and high density, the quarks 
form Cooper pairs, associated with a Higgs 
field $\phi^{\al\be}_{ij} \sim \psi^\al_i\psi^\be_j$.
This field 
gets a vacuum expectation value which takes the form
\begin{equation}
\langle \phi^{\al\be}_{ij}\rangle \sim \Delta \epsilon^{\al\be} \epsilon_{ij}
\end{equation}
in the 2SC phase, with $i$ and $j$ running over $u$ and $d$ only and
$\al$ and $\be$ running over two of the colors only.
In the CFL phase, quarks of all three flavors and colors
pair and the expectation value of the field takes the form
\begin{equation}
\langle \phi^{\al\be}_{ij}\rangle \sim 
\Delta_1 \delta^\al_i\delta^\be_j + \Delta_2 \delta^\al_j\delta^\be_i\ ,
\end{equation}
with all indices now taking on three values.
In both the CFL and 2SC phases, the condensate leaves
an unbroken $U(1)$ generated
by
\beql{rot:Qprime}
\R = Q + \eta \Y, \qquad \R\<\phi\> = 0,
\eeql
where $\eta=1/\sqrt{3}$ for CFL, and $\eta=-1/(2\sqrt{3})$ for 2SC.
$Q$ is the conventional electromagnetic charge generator,
and $\Y$ is associated with one of the gluons.
In the representation of the quarks,
\beq
\ba{rcl@{\,\,}ll}
Q &=& &\mbox{diag}(\twothirds,-\third,-\third) &
\mbox{in flavor $u,d,s$ space} \\[1ex]
\Y &=& \frac{1}{\sqrt{3}} & \mbox{diag}(1,1,-2) &
\mbox{in color $r,g,b$ space}. \\
\ea
\eeq
As is conventional, we have taken $\tr(\Y \Y) = 2$.
The 
$\R$-charge of all the Cooper pairs which form the condensates
are zero.

To see exactly 
which gauge field remains unbroken, look at the covariant
derivative of the Higgs field:
\beq
D_\mu\<\phi\> = 
\Bigl(\p_\mu + eA_\mu Q^{(\phi)} + g {G^8_\mu}\Y^{(\phi)}\Bigr)\<\phi\>
\eeq
{}From \eqn{rot:Qprime}, we see that the kinetic term $|D\<\phi\>|^2$
will give a mass to one gauge field
\beq
\As_\mu = \frac{-\eta e A_\mu + g G^8_\mu}{\sqrt{ \eta^2 e^2 + g^2}}
= -\sin\al_0 A_\mu + \cos\al_0 G^8_\mu
\eeq
but the orthogonal linear combination 
\beql{rot:Aprime}
\Ar_\mu = \frac{g A_\mu + \eta e G^8_\mu}{\sqrt{ \eta^2 e^2 + g^2}}
= \cos\al_0 A_\mu + \sin\al_0 G^8_\mu
\eeql
will remain massless.
The denominators arise from keeping the gauge field kinetic terms
correctly normalized, and we have defined the 
angle $\al_0$,
\beql{rot:alpha0}
\cos\al_0 = \frac{g}{\sqrt{ \eta^2 e^2 + g^2}}\ ,
\eeql
which specifies the unbroken $U(1)$.
At neutron star densities the gluons are strongly coupled
($g^2/(4\pi) \sim 1$), 
and of course the photons are weakly coupled
($e^2/(4\pi) \approx 1/137$), so $\al_0\simeq \eta e/g$ 
is small: $\alpha_0\sim 1/20$ in the 2SC phase and 
$\alpha_0\sim 1/40$ in the CFL phase.
In both phases, the ``rotated photon''
consists mostly of the usual photon, with only a small
admixture of the $\Y$ gluon.

We can now see that the electron couples to $\Ar$ with charge
$ e g /\sqrt{ \eta^2 e^2 + g^2}$ which is less than $e$, so
the new photon is slightly more weakly coupled to electrons 
than the old one.  It can also be shown that in the CFL phase, the quarks
have $\R$-charges which are integer 
multiples of the $\R$-charge of the electron \cite{ARW3}.

\subsection{The Meissner effect}
The main purpose of this paper is to study the Meissner effect in a
high density region, which we will assume to be a sphere, like the
core of a neutron star, with radius $R$. In this region, color
superconductivity occurs.  The unbroken generator $\R$ is associated
with a rotated electromagnetic field $A^\R_\mu$, 
with rotation angle $\al_0$, as described above.
The broken generator $\S$ gives no long-range gauge field.  
For simplicity we will treat the region outside the core not as
nuclear matter, but as vacuum. As we will explain in
Sect.~\ref{sec:outlook}, this does not seriously affect our
conclusions. In the external region, then, ordinary
electromagnetism $A^Q_\mu$ is unbroken, and the color generator $\Y$
is confined.  We assume that currents at infinity create a uniform
applied $Q$-magnetic field $B^Q_a$ in the $z$-direction.  We want to
know to what degree the flux is expelled from the inner region.

To study this situation, we need only work in the two-dimensional
space of gauge symmetry generators spanned by $Q$ and $\Y$. It is
natural to take $\al$ (the angle by which the unbroken $U(1)$ is
rotated relative to ordinary electromagnetism, see \eqn{rot:alpha0})
to vary with radius, with $\al(r)=\al_0$ in the high density
(inner) region, and $\al(r)=0$ outside. So in the $(Q,\Y)$ basis,
\beql{rot:basis}
 Q = \left(\ba{@{}c@{}}1 \\ 0 \ea \right),\quad
\Y = \left(\ba{@{}c@{}}0 \\ 1 \ea \right),\quad
\R =  \left(\ba{@{}c@{}}\cos\al(r) \\ \sin\al(r) \ea \right),\quad
\S =  \left(\ba{@{}c@{}} -\sin\al(r) \\ \phm\cos\al(r) \ea \right).
\eeql

To determine what fraction of the magnetic field is
expelled from the color superconducting region, we must consider
what happens at the boundary between the two phases. 
There are three relevant distance scales:
\beql{rot:scales}
\ba{ll}
R & \mbox{the size of the high-density region}, \\[1ex]
\de R\sim (d\al/dr)^{-1} & 
  \parbox{3.5in}{the distance over which $\al$
          switches from $\al_0$ to 0, the boundary thickness} \\[2ex]
\la & \mbox{the screening distance for broken/confined gauge fields}
\ea
\eeql

We will always assume that $R \gg \la$. We will treat two
limiting scenarios: ``sharp'' boundary and ``smooth'' boundary.
The sharp boundary, $\de R \ll \la$, corresponds to a sudden
step-function-like change in $\al$. This is what we would
expect to find if there were a first-order phase transition
between the low and high density regions, with phase boundary thickness
less than the screening length.
The smooth boundary, $\de R \gg \la$, corresponds to a gradual
change in $\al$. This applies to the situation where there is no
first-order phase transition between nuclear and quark matter 
(e.g.~for low strange quark mass,
where we expect no phase transition at all \cite{SchaeferWilczek,ABR,SW2}),
or where there is a first-order phase transition with 
phase boundary thicker than the screening length.
We will see that the behavior of the magnetic field is quite different
depending on whether the boundary is sharp or smooth.

What we are interested in is the behavior of the magnetic field at
macroscopic distance scales of order $R$, not at the microscopic
scale $\la$. In other words, we will always be interested in
the unbroken and unconfined gauge fields, which obey the Maxwell
equations.  In order to treat all gauge fields in a single formalism
it will therefore be convenient to take into account
screening not by introducing gauge boson masses into the field
equations, but rather by introducing
monopoles and supercurrents to describe the screening of the confined
and Higgsed gauge fields.  By ``monopoles'' we mean whatever gauge field
configurations are responsible for terminating magnetic field lines
of confined gauge fields; we do not require them to be solutions
to any classical field equations.
We will ignore the details of how the
supercurrents (and ``monopoles'') arrange themselves on distance scales
of order $\la$, as all we care about is that they screen (terminate)
the higgsed (confined) magnetic fields incident upon them.  

\section{Sharp boundary}
\label{sec:sharp}

\begin{figure}[thb]
\begin{center}
\epsfig{file=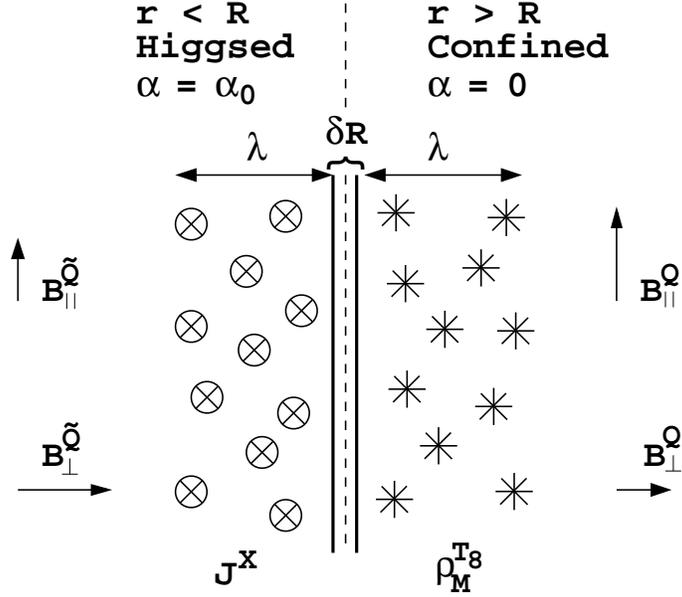,width=3.5in}
\end{center}
\caption{Cross-section of a small region of the sharp
boundary, showing the location of the $\S$-supercurrents 
(which ensure that there is no parallel $X$-flux inside
the color superconducting sphere at $r\lesssim R-\la$)
and the $\Y$-monopoles (which ensure that there is no
perpendicular $T_8$-flux outside the sphere at 
$r\gtrsim R+\la$.)
}
\label{fig:sharp}
\end{figure}
In the sharp boundary case, $\al$ changes quickly from $\al_0$ to 0
at $r=R$ over a boundary region with
thickness $\delta R < \la$ (Fig.~\ref{fig:sharp}). 
Just inside the boundary, in the region $R-\la \lesssim r < R$, 
the $\S$ gauge field is Higgsed, so there
is a density $\vec J^\S$ of $\S$-supercurrents that screen out the
parallel component of the 
$\S$-flux, leaving only $\R$-flux at $r\lesssim R-\la$.
Just outside the boundary, in the region $R<r\lesssim R+\la$,
the $\Y$ gauge field is confined, so
there is a density $\rho^{\Y}_M$ of $\Y$-monopoles that terminate
any perpendicular component of the $\Y$-flux, 
leaving only $Q$-flux at $r\gtrsim R+\la$.

To obtain the boundary conditions on the magnetic fields
associated with the unbroken generators, 
write the magnetic field  in the $(Q,\Y)$
basis \eqn{rot:basis}, 
\beql{sharp:basis}
\vec B = \colvec{\vec B^Q}{\vec B^{\Y}} = 
 \colvec{\cos\al_0\, \vec \Br - \sin\al_0\, \vec \Bs}{\sin\al_0\, \vec \Br + \cos\al_0\, \vec \Bs},
\eeql
then the Maxwell equations are
\beql{sharp:maxwell}
\ba{rrclrcl}
r>R:~ & \div\,\vec B  &=&  \colvec{0}{\rho^{\Y}_M},\quad
 & \curl\,\vec B &=& 0,\\[1.5ex]
r<R:~ & \div\,\vec B &=& 0,\quad
 & \curl\,\vec B &=&  \colvec{-\sin\al_0\, \vec J^\S}{\phm\cos\al_0\, \vec J^\S}.
\ea
\eeql

%

Since we are only interested in the behavior of the fields on distance
scales much greater than $\la$, we integrate the Maxwell equations
\eqn{sharp:maxwell} over $R-\la < r < R+\la$, and obtain
boundary conditions that relate the fields at $R-\la$ to
those at $R+\la$.
We follow the standard derivation (Ref.~\cite{Jackson}, sect.~I.5).
We find a discontinuity in 
the normal compoment $B^{\Y}_\bot$ due to the surface density of
$\Y$-monopoles, and in 
the parallel component $\Bs_\|$ due to surface $\S$-supercurrents,
%
%
\beql{sharp:discontinuity}
\ba{rcl}
B^Q_\bot(R\!+\!\la) \colvec{1}{0} 
  - \Br_\bot(R\!-\!\la)\colvec{\cos\al_0}{\sin\al_0}
  &=& \rho^{\Y}_M \colvec{0}{1}, \\[2ex]
B^Q_\|(R\!+\!\la) \colvec{1}{0} 
  - \Br_\|(R\!-\!\la)\colvec{\cos\al_0}{\sin\al_0}
  &=& J^\S \colvec{-\sin\al_0}{\phm\cos\al_0}.
\ea
\eeql
{}From this we immediately obtain the boundary conditions on the flux:
\beql{sharp:bc}
\ba{rcl}
\Br_\bot(R\!-\!\la) &=& \dsp \frac{1}{\cos\al_0}\, B^Q_\bot(R\!+\!\la), \\[2ex]
\Br_\|(R\!-\!\la) &=& \cos\al_0\, B^Q_\|(R\!+\!\la).
\ea
\eeql
Thus, just inside the interface we find a $\vec B^{\R}$ whose
component parallel (perpendicular) to the interface is 
weakened (strengthened) relative to that
of $\vec B^{Q}$ just outside the interface.

\subsection{Solution for the sharp boundary}
\begin{figure}[htb]
\begin{center}
\epsfig{file=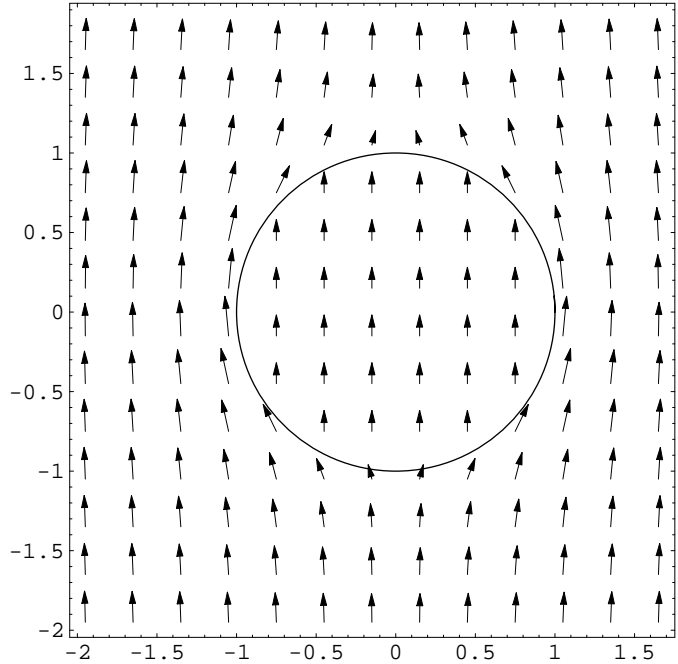,width=3.5in}
\end{center}
\caption{
Locally unbroken magnetic field inside and outside a sphere
of high-density quark matter, with sharp boundary.
This is \eqn{sharp:solution}, with $\cos\,\al_0 = 0.5$.
}
\label{fig:fields}
\end{figure}

Since this is a magnetostatic problem we can write it in terms of
a magnetic scalar potential $\Phi$. The potential is associated with the
unbroken $\R$ flux inside the sphere, and the unbroken $Q$-flux outside
\beq
\ba{rclr}
B^Q &=& - \nabla \Phi & \mbox{outside sphere}, \\
\Br&=&  - \nabla \Phi & \mbox{inside sphere}.
\ea
\eeq
Maxwell's equations become
\beq
\nabla^2 \Phi = 0,
\eeq
with boundary conditions \eqn{sharp:bc} 
\beq
\ba{rcl}
\Phi(r\to\infty) &=& B^Q_a r \cos\th, \\[1ex]
\dsp \frac{\p \Phi}{\p r}(R\!-\!\la) &=& \dsp 
  \frac{1}{\cos\al_0} \frac{\p \Phi}{\p r}(R\!+\!\la), \\[3ex]
\dsp \frac{\p \Phi}{\p \th}(R\!-\!\la) &=& \dsp 
  \cos\al_0 \frac{\p \Phi}{\p \th}(R\!+\!\la).
\ea
\eeq
Expanding in Legendre polynomials (see \cite{Jackson} sect.~5.12)
we find the solution
\beql{sharp:solution}
\ba{rcll}
\Phi &=& \dsp -B^Q_a r \cos\th  - B^Q_a\frac{R^3}{r^2}\cos\th\,\,
 \frac{1 - \cos^2\al_0}{2+\cos^2\al_0} 
 & (r>R), \\[3ex]
\Phi &=& \dsp -B^Q_a r \cos\th \,\, \frac{3\cos\al_0}{2+\cos^2\al_0}
 & (r<R).
\ea
\eeql
In Fig.~\ref{fig:fields} we show the resultant field configuration for
$\cos \al_0=0.5$.
In the real world $\al_0$ is small, so the
field is mostly converted into $\R$ flux by the supercurrents and
monopoles, and penetrates
the interior. Only a weak field is excluded.

We can check that the solution \eqn{sharp:solution} makes sense
in two limits.
(1) $\al_0=0$, so $\R=Q$ and $\Y=\S$. The unbroken $U(1)$ in the
color superconducting region
is exactly the same as conventional electromagnetism: the magnetic
field does not even notice the color superconductor.
(2) $\al_0=\pi/2$, so $\R=\Y$ and $\S=Q$. The region is a conventional
superconductor, breaking electromagnetism. The magnetic field is 
completely expelled from the superconducting region. 

How large a field must we apply before we begin to destroy
the color superconductivity in parts of the sphere?
The magnitude of the $Q$-magnetic field just outside the sphere
is largest at the equator of the sphere, where it is 
given by
\begin{equation}
|\vec B^Q(r=R+\lambda,\theta=\pi/2)| = B_a^Q \left( 1+
\frac{1-\cos^2\alpha_0}{2+\cos^2\alpha_0}\right) \ \label{demag}.
\end{equation}
This means that the equator of the sphere sees an $X$-magnetic field
given by $\sin \alpha_0$ times (\ref{demag}).  If the color
superconductor is Type I, the color
superconductivity is destroyed in regions of the sphere near its
equator when the $X$-magnetic field exceeds the thermodynamic
field $H_c\sim 10^{18}$ Gauss.\footnote{If the material is Type II, flux tubes
begin to penetrate when the $X$-magnetic field exceeds $H_c^1$,
which is of the same order of magnitude.} This
requires
\begin{equation}
B_a^Q > \frac{2+\cos^2\alpha_0}{3 \sin \alpha_0} H_c \ .
\end{equation}
We conclude that because most of the applied field can penetrate
the superconductor in the form of $\tilde Q$-flux, while only
a small fraction of the applied field
must be excluded, the applied field at which
the color superconductivity begins to be destroyed
is significantly larger than the thermodynamic critical
field $H_c\sim 10^{18}$ Gauss.

\section{Smooth boundary}
\label{sec:smooth}

For a smooth boundary, $\de R \gg \la$, which means that $\al$ changes
slowly relative to the screening length $\la$. In this case we assume
that there is a continuous distribution of monopoles and supercurrents.
At a given $r$ these produce flux in the $U(1)$ that is
(locally) broken.  Since the screening length $\la$ is small, the net
flux at a given $r$ must only be in the unbroken $U(1)$.  The Maxwell
equations therefore take the form
\beql{smooth:maxwell}
\ba{rcl}
 \div\,\left\{ \vec B\colvec{\cos\al(r)}{\sin\al(r)} \right\}
  &=& \rho_M \colvec{-\sin\al(r)}{\phm\cos\al(r)}, \\[2ex]
 \curl\,\left\{ \vec B\colvec{\cos\al(r)}{\sin\al(r)} \right\}
  &=& \,\,\,\vec J \colvec{-\sin\al(r)}{\phm\cos\al(r)},
\ea
\eeql
where $\vec B,\rho_M,\vec J$ are functions of position.
Now
\beq
\ba{rcl}
\div \left\{ \vec B\colvec{\cos\al(r)}{\sin\al(r)} \right\}
&=& \dsp \div\vec B \colvec{\cos\al(r)}{\sin\al(r)}
+ B_r \frac{d \al}{dr} \colvec{-\sin\al(r)}{\phm\cos\al(r)}, \\[3ex]
\curl \left\{ \vec B\colvec{\cos\al(r)}{\sin\al(r)} \right\}
&=& \curl\vec B \colvec{\cos\al(r)}{\sin\al(r)}
+ \vec B \times \del\al \colvec{-\sin\al(r)}{\phm\cos\al(r)}.
\ea
\eeq
Substituting these into \eqn{smooth:maxwell} and taking the
$(\cos\al, \sin\al)$ component, we find that the locally unbroken
magnetic field $\vec B$ obeys the sourceless Maxwell equations:
\beql{smooth:free}
\div\vec B = 0, \qquad \curl\vec B = 0.
\eeql
We conclude that in this case the magnetic field always rotates
to be locally unbroken, but is otherwise unaffected: it obeys
the free Maxwell equations {\em everywhere}. There is no expulsion
at all. (In the smooth boundary case, then,
the effective critical magnetic field
is infinite: an arbitrarily strong magnetic field is always
rotated to be in the locally unbroken $U(1)$, and never
destroys the color superconductivity.)

One way of understanding this result is to note that the smallness of
the screening length $\la$ means that broken gauge fields are
always zero. Consequently, when we move from radius $r$
to $r+dr$, the gauge field is immediately projected onto the
locally unbroken generator $\R(r+dr)$, which differs by an
infinitesimal angle from $\R(r)$. As we vary $r$, we therefore have
a sequence of such projections. In the limit of an infinite number
of projections, each infinitesimally different from the last,
the gauge field is simply rotated to follow the locally unbroken generator,
\beq
\lim_{N\to\infty} \prod_{n=1}^N \cos(\frac{\al_0}{N}) = 1.
\eeq
This is analogous to the behavior of a
sequence of $N$ polarizers each at an angle
$\al_0/N$ to the previous one. In the limit where $N\to\infty$,
an incoming photon aligned with the first polarizer
exits from the last polarizer with its polarization rotated by $\al_0$,
and with no loss of intensity.

Finally, it is interesting to ask why the $\de R\to 0$ limit of our
result for the smooth boundary is not the same as our result for the
sharp boundary. The reason is that in the smooth case we assume
$\la\ll\de R$, so the monopoles and supercurrents are always in the
region where $\al$ is changing. In the sharp case, by contrast, we
assume $\de R\ll \la$, and there is in effect no such region.
Taking the $\de R \rightarrow 0$ limit of the smooth case
($\la\ll\de R$) would result in a pile-up of currents and monopoles in
a shrinking region within which $\alpha$ changes rapidly from $\al_0$
to 0.  In this limit, (\ref{smooth:free}) is maintained and no flux is
excluded from the sphere of color 
superconductor.\footnote{Formally, taking the   $\de R \rightarrow 0$
limit while maintaining $\la\ll\de R$ yields both monopoles and
supercurrents concentrated at a point where $\alpha=\alpha_0/2$.
If one modifies the right hand side of (\ref{sharp:discontinuity})
to reflect this, the boundary conditions (\ref{sharp:bc})
becomes $\vec B^Q(R\!-\!\la)=\vec B^Q(R\!+\!\la)$, and
no flux is excluded.}
The sharp boundary studied in Sect.~\ref{sec:sharp} represents
different physics, in which $\la$ is constant as $\de R\to 0$.  In
this limit all the monopoles are in the $\al=0$ region on the confined
side of the boundary while all the supercurrents are in the
$\al=\al_0$ region on the Higgsed side and we obtain 
the boundary condition (\ref{sharp:bc}) and 
partial flux
exclusion as in (\ref{sharp:solution}).

\section{Consequences for Neutron Stars and Outlook}
\label{sec:outlook}

If a neutron star features a core made of color superconducting
quark matter, we have learned that this 
core exhibits (almost) no Meissner
effect in response to an applied magnetic field.  Even though
the Cooper pairs of quarks have nonzero electric charge,
there is an unbroken gauge symmetry $U(1)_{\tilde Q}$,
and the color superconducting region can support a $\tilde Q$-magnetic 
field. If the
boundary layer of the color superconducting region is
thick compared to the screening length, there is {\it no}
Meissner effect: the $\tilde Q$-magnetic field within
has the same strength as the applied $Q$-magnetic field.
(This smooth boundary case applies if the color 
superconducting phase and the baryonic phase are not separated
by a phase transition; it may also apply  in the presence
of a first order phase transition, if the phase boundary
is thick enough.)
If the thickness of the boundary of the superconducting
region is less than the penetration length, there is
a partial Meissner effect: the $\tilde Q$-magnetic field
within is somewhat reduced relative to the applied $Q$-magnetic
field.  Because $\alpha_0$ is small in nature, the overlap
between $Q$-photons and $\tilde Q$-photons is large, 
and this partial Meissner effect occurs only at the few
percent level.  

The physics of magnetic fields in neutron stars with 
color superconducting cores is qualitatively
different from that in conventional neutron stars.  
In conventional neutron stars, proton-proton pairing
breaks $U(1)_Q$ and there is no unbroken $U(1)_{\tilde Q}$.
This means that magnetic fields thread the cores of conventional
neutron stars in quantized flux tubes, within each of which
there is a nonsuperconducting region.  
In contrast, the $\tilde Q$-magnetic
field within a color superconducting neutron star core 
is not confined to flux tubes.  This means
that, as we discuss below, the enormous 
$\tilde Q$-electrical conductivity of the matter ensures
that the $\tilde Q$-magnetic field is constant in time.
In ordinary neutron stars the $Q$-magnetic
flux tubes can be dragged about by the outward motion of the rotational
vortices as the neutron star spins down  
\cite{Sauls,Dragging,Bhattacharya,Ruderman,RudermanTalk},
and can also be pushed outward if the gap at the 
proton fermi surface increases with depth within
the neutron star \cite{HsuMag}.
One therefore expects the magnetic field of an isolated pulsar to
decay over billions of years as it spins down
\cite{Dragging,Bhattacharya,Ruderman,RudermanTalk}
or perhaps more quickly \cite{HsuMag}.  
There is
no observational evidence for the decay of the magnetic
field of an isolated pular over periods of billions of
years \cite{Bhattacharya}; this is consistent with the
hypothesis that they contain color superconducting cores
which serve as ``anchors'' for magnetic field, because
they support a $\tilde Q$-magnetic field which does
not decay.

We now estimate the decay time of the $\tilde Q$-magnetic
field for neutron stars with color superconducting cores,
doing the calculation separately for
cores which are in the 2SC and the CFL phase.
To this point, the only difference between
the two color superconducting phases in their response
to applied magnetic fields has been a difference of
a factor of two in the value of $\alpha_0$; because
$\alpha_0\ll 1$ in both phases, this difference is
of no qualitative consequence.  
However, the 2SC and the CFL phase differ qualitatively in their 
symmetries and their low-energy excitations, and can therefore
be expected to have quite distinct transport properties.
As we will see below,
the $\tilde Q$-electrical conductivity in the CFL phase is much
larger than the conductivity in the CFL phase. However, even the 
``smaller'' conductivity of the 2SC phase is so large that
the timescale for the decay of a 
$\tilde Q$-magnetic field within a color superconducting
neutron star core is long 
compared to the age of the universe.

The characteristic magnetic field
decay time due to ohmic dissipation is \cite{BPP}
\begin{equation}
t_{\rm decay}\sim 4\sigma R^2/\pi \ ,
\end{equation}
where $R$ is the radius of the color superconducting core
and $\sigma$ is the $\tilde Q$-electric conductivity. 
(We set $\hbar=c=k_B=1$ throughout.) 
We begin by estimating $\sigma$, and hence the 
decay time, for a core 
which is in the 2SC phase. 
At keV temperatures, the 
dominant carriers are the relativistic electrons and 
those quarks of the 2SC phase which are ungapped, or which acquire gaps
so small as to be $\sim T$ or less.  In the 2SC phase 
the up and down quarks of one color acquire gaps smaller than
or of order keV \cite{ARW2}.  The strange quarks 
of all three colors --- which do not participate in
the dominant pairing characterizing the 2SC phase --- 
can be expected to have gaps which are similar in magnitude
or even smaller \cite{ABR}.\footnote{
The contribution from the quark quasiparticles with 
gaps $\Delta \sim 20-100 {\rm MeV} \gg T$ can be neglected.
In the CFL phase, {\it all} quark quasiparticles 
have gaps which are $\gg T$. We discuss this situation
below.}
To obtain a lower bound on $\sigma$, we assume that 
all strange quarks, and up and down quarks of one color,
have gaps $\ll T$. These five quarks have $\tilde Q$-charges
$-\half,-\half,0,1,0$ in units of the $\tilde Q$-charge
of the electron $eg/\sqrt{g^2+\e^2/12}$, which we henceforth
take to be just $e$ since $e/g\ll 1$.  The 
sum of the squares of the $\tilde Q$-charge
of the ungapped quarks is therefore $3e^2/2$.
Following \cite{HP}, 
the electrical conductivity is given by
\begin{equation}
\sigma \sim \frac{1}{3\pi^2} \mu_e^2 e^2 \tau_e 
+ \frac{3}{2}\,\frac{1}{3\pi^2}
\mu^2 e^2 \tau_q\  ,
\end{equation} 
where $\tau_e$ and $\tau_q$ are the momentum relaxation times
for electrons and quarks in the plasma, defined and 
calculated in Ref. \cite{HP}.
For both the electrons and quarks, the dominant scattering
process contributing to momentum relaxation is  
scattering off quarks. 
The five gapless quarks of interest yield \cite{HP}
\begin{equation} 
\tau_q^{-1} \simeq \frac{40}{9\pi}\, 1.81\, \alpha_s^2 \,
\frac{T^{5/3}}{(m_D^g)^{2/3}} 
\label{tauqeq}\end{equation}
where $\alpha_s=g^2/4\pi$ and $(m_D^g)^2 \simeq 3 g^2 \mu^2 /(2\pi^2)$
is the Debye screening mass for the gluons,
neglecting $M_s$ relative to $\mu$,
and where we have assumed that $T\ll m_D^g$.\footnote{Note
that we have worked directly from
equations (28) and (39) of Ref. \cite{HP}, 
and have not used the numerical factor
in equation (40) of Ref. \cite{HP}, 
which contains an error \cite{HPprivate}.} 
Similarly,
\begin{equation} 
\tau_e^{-1} \simeq \frac{4}{3\pi}\, 1.81\, \alpha_e^2 \,
\frac{T^{5/3}}{(m_D^e)^{2/3}}\ , 
\label{taueq}\end{equation}
where $\alpha_e=1/137$, and $(m_D^e)^2 \simeq 5 e^2 \mu^2 /(12 \pi^2)$
is the Debye screening mass for the $\tilde Q$-photon.
We have neglected $M_s$, used the fact that the average squared $\R$-charge
of the nine quarks participating in screening is $5e^2/18$, and have
assumed $T\ll m^e_D$.
Taking $\al_s\sim 1$, we find first that the
electrons dominate the conductivity, by a factor of about $20$,
and second that the
time-scale for the decay
of the magnetic field in a color superconducting 2SC core
of radius $R$ is 
\begin{equation}
t_{\rm decay}\sim 3\times 10^{13}\,  
{\rm years} \ \left(\frac{R}{1\, {\rm km}}\right)^2
\left(\frac{\mu}{400\, {\rm MeV}}\right)^{2/3} 
\left(\frac{\mu_e}{25\, {\rm MeV}}\right)^2
\left(\frac{T}{1\, {\rm keV}}\right)^{-5/3} \ .
\label{decaytime}
\end{equation}
Thus, the magnetic field in the core of a neutron star
which is made of matter in the 2SC phase decays only
on a time-scale which
exceeds the age of the universe. 

We now turn to a color superconducting core in the CFL phase. 
In contrast to the 2SC phase, the condensation in the
CFL phase produces gaps $\gg T$ for quarks of all three colors 
and all three flavors.\footnote{It may be possible for some 
quark quasiparticles to be gapless even in the CFL phase, at densities just
above those where the 2SC phase is favored \cite{gapless}. In
this nongeneric circumstance, the CFL phase conductivity would
be similar to that of the 2SC phase.}
Thus there are {\em no} low-energy quark quasiparticle excitations.
Furthermore, the CFL condensate gives a mass to all eight gluons.
In this quark matter phase, the degrees of freedom
which are most easily excited  are neither
quarks nor gluons.
Because of the spontaneous breakdown
of chiral symmetry in the CFL phase, there are  
charged Nambu-Goldstone bosons, which would
be massless pseudoscalar excitations
if the quarks were massless. Once the nonzero
quark masses are taken into account, one finds
pseudoscalar masses which are small \cite{RappEtc2} 
(in the sense that they are $\ll \Delta$)
but which are still large compared to $T$.
There is one remaining scalar Nambu-Goldstone excitation,
associated with the superfluidity of the CFL phase, 
but this excitation is $\tilde Q$-neutral.
We thus discover that
all possible hadronic excitations which have nonzero $\tilde Q$-charge
{\em do} acquire a gap:  their populations in thermal equilibrium
are suppressed exponentially by factors of the form $\exp(-\Delta/T)$,
where $\Delta$ is either a Fermi surface gap or a suitable
bosonic mass.
This first of all means that the only charge carriers which
could contribute to $\sigma$ are the electrons. Second, 
the scattering of these electrons off the positively charged
hadronic system in which they are immersed vanishes exponentially
for $T\rightarrow 0$, because at low temperatures there are
no hadronic excitations off which to scatter.\footnote{We 
are describing the conductivity in the linear response regime.
In particular, we are assuming that the current
is small enough that the momenta acquired by the electrons
due to the current is small compared 
to the excitation energy for all charged hadronic 
modes. If the current were increased beyond
the linear regime, the conductivity
would decrease and eventually the CFL condensate itself would
be destroyed.  The nonlinear 
regime is not relevant for our purposes.} 
The electrons can only scatter off other electrons. However, 
such collisions do not alter the total electric
current \cite{Conductivity}. 
We therefore conclude that although matter in the CFL phase is not a
$\tilde Q$-superconductor (it does not exclude $\tilde Q$-magnetic
field) it is a near-perfect conductor: the resistivity drops
exponentially to zero as $T\to 0$. At typical neutron star
temperatures $T\ll \Delta$, the density of hadronic excitations, and therefore 
the resistivity, is exponentially 
suppressed relative to the 2SC phase. As a consequence, the decay time
for a $\tilde Q$-magnetic field in a CFL core is exponentially larger
than that (\ref{decaytime}) for a core in the 2SC phase, which was
already longer than the age of the universe.

It is clear that the $\tilde Q$-magnetic field is rigidly
locked in the color superconducting core. It cannot decay
with time, even if rotational vortices
move through the core as the spin rate of the pulsar
changes with time.  Rotational vortices
do exist in the CFL phase, because the CFL condensate 
spontaneously breaks a global $U(1)$. 
Instead of dragging magnetic
flux tubes with them as they move, as occurs in 
a conventional neutron star, the rotational vortices can move 
freely through the CFL phase because there are no
$\tilde Q$-flux tubes, only a $\tilde Q$-magnetic field. 
Thus, as the spin period of the neutron star changes
and the rotational vortices move accordingly, 
there is no change at all in the strength of the 
$\tilde Q$-magnetic field in the core.
The conclusion is the same in the 2SC phase, but the
argument is even simpler because in this phase there is no
spontaneously broken global $U(1)$, and therefore no
rotational vortices.

As we have noted above, the data on isolated pulsars
show no evidence for any decay of the observed magnetic 
field even as they spin-down over
time \cite{Bhattacharya}.  This is consistent with
the possibility of a color superconducting core 
within which the field does not decay.
However, we must also ask whether and how the fact
that the observed magnetic fields of accreting pulsars {\it do} 
change as they spin up \cite{Bhattacharya}
is consistent with the possibility of quark matter cores.
There are several ways in which the surface magnetic field
could decrease as a neutron star accretes and spins up, even
though the magnetic field in the core remains constant. One
possibility is that the accreting matter may bury
the magnetic field \cite{FieldBurial,Bhattacharya}.
Another possibility is that as the magnetic flux tubes
in the mantle and crust of the neutron star are pulled around
by the rotational vortices, 
the north and south magnetic poles on the star's surface
may be pushed toward
one another, reducing the observed dipole field 
even though the field deep within, in
the color superconducting
core, remains undiminished \cite{Ruderman,RudermanTalk}.

The analysis of Section \ref{sec:sharp} was idealized in three ways
relative to that appropriate for a neutron star, if there is a
first order phase transition between baryonic and quark matter. First,
we assumed that ordinary electromagnetism was unbroken outside the
core.  This is false: proton-proton superconductivity results in the
restriction of ordinary $Q$-magnetic field to flux tubes.  This means
that our derivation of the boundary conditions in Section \ref{sec:sharp} only
applies upon averaging over an area of the boundary that is
sufficiently large that many $Q$-flux tubes are incident on it from
the outside.  A more microscopic description of the field
configuration near the boundary would in fact require further work,
but this is not necessary for our purposes.

Second, our assumption in Section \ref{sec:sharp}
of a spherical boundary is oversimplified. 
If there is a first order phase transition between
baryonic and quark matter, because there are two distinct
chemical potentials $\mu$ and $\mu_e$ there will be a mixed
phase region, with many boundaries separating regions of
quark matter and baryonic matter which have complex shapes \cite{Glendenning}.
This complication of the geometry of the boundary evidently makes
a complete calculation more difficult than the one we have
done in Section \ref{sec:sharp}, although if the boundary is thick
then the conclusions of Section \ref{sec:smooth} are unaffected. 
Regardless, the qualitative conclusion that only a very small fraction
of the flux will be excluded because $\alpha_0$ is so small
will not be affected by these complications.

Third, the configuration of Figure 2 cannot, in fact, be attained
in a neutron star even though it is favored in the sharp boundary
case.
The core of 
a newborn neutron star is threaded with ordinary $Q$-magnetic field. 
Because of the enormous conductivity, the
time it would take to accomplish the partial exclusion of 
flux seen in Figure 2 is exceedingly long \cite{BPP}.
This means that, instead, although the field within
the core will be largely $\tilde Q$-magnetic
field, there will in addition be a small fraction
of $X$-magnetic flux confined in quantized flux tubes. The 
sum of the $\tilde Q$- and $X$-fluxes adds up to the original
$Q$-flux.  Over time, the motion of rotational vortices may 
move the $X$-flux tubes around.  The much larger 
$\tilde Q$-flux, which
is not constrained in flux tubes, is frozen as described above.

We conclude that relaxing the simplifying assumptions
which we have made would not change our qualitative
conclusions.
If neutron stars contain quark matter cores,
those cores will exclude at most a very small fraction of
any applied magnetic field. Instead, the flux penetrates
(almost) undiminished.  The only
change in the flux within the color superconducting core
is that it is a $\tilde Q$-magnetic field, associated with that linear
combination of the photon and the $T_8$ gluon which is
unbroken by the quark pair condensate.  Most important
for neutron star phenomenology, and in qualitative distinction
from the results for conventional neutron stars,
is the conclusion that this
$\tilde Q$-flux does not form quantized flux tubes and is 
frozen over timescales long compared to the age of the
universe.

We are confident that we have not said the last
word on the effects of color superconducting
quark matter cores within neutron stars on
magnetic field evolution. 
More work is required to better understand pulsars
which are accreting and spinning up.
We can already conclude that
if observational evidence were to emerge 
that an {\em isolated} pulsar loses its magnetic field
as it spins down (in such a way that the field would 
vanish in the limit in which the spin vanishes),
this would allow one to infer that such a pulsar
does not have a quark matter core.

\vspace{3ex}
{\samepage 
\begin{center} Acknowledgements \end{center}
\nopagebreak
We are grateful to the Aspen Center for Physics, where much of
this work was completed.
We thank I. Appenzeller,
M. Camenzind, D. Chakrabarty, H. Heiselberg, L. Hernquist,
R. Jaffe, V. Kaspi, D. Psaltis, M. Ruderman, T. Sch\"afer, 
E. Shuster and F. Wilczek for helpful discussions.
This work is supported in part  by the U.S. Department
of Energy (D.O.E.) under cooperative research agreement \#DF-FC02-94ER40818.
The work of KR is supported in part by a DOE OJI Award and by the
Alfred P. Sloan Foundation.
}

\end{document}